\documentclass[a4paper,10pt,twoside]{cpc-hepnp}

\usepackage{multicol}
\usepackage{graphicx}
\usepackage{booktabs}
\usepackage{amssymb,bm,mathrsfs,bbm,amscd}
\usepackage[tbtags]{amsmath}
\usepackage{lastpage}

\begin{document}

\fancyhead[co]{\footnotesize CHEN Guo-Chang~ et al: Neutron Nuclear Data Evaluation of Actinoid Nuclei for CENDL-3.1}

\footnotetext[0]{Received }

\title{Neutron Nuclear Data Evaluation of Actinoid Nuclei for CENDL-3.1}

\author{%
      CHEN Guo-Chang$^{1;1)}$\email{cgc@ciae.ac.cn}%
\quad CAO Wen-Tian$^{2}$%
\quad YU Bao-Sheng$^{1}$\\
\quad TANG Guo-You$^{2}$
\quad SHI Zhao-Min$^{2}$
\quad TAO Xi$^{1}$
}
\maketitle

\address{%
$^1$ Laboratory of Science and Technology on Nuclear Data, China Institute of Atomic Energy, Beijing 102413, China;\\
$^2$ School of Physics $\&$ State Key Laboratory of Nuclear Physics and Technology, Peking University, Beijing 100871, China\\
}

\begin{abstract}
New evaluations for several actinoids of the third version of China Evaluated Nuclear
Data Library (CENDL-3.1) have been completed during the period between 2000 and 2005.
The evaluations are for all neutron induced reactions with Uranium, Neptunium,
Plutonium and Americium in the mass range A=232-241, 236-239, 236-246 and 240-244, respectively,
and cover the incident neutron energy up to 20 MeV. In present evaluation, much more efforts
were devoted to improve reliability of nuclide for available new measured data, especially scarce
experimental data. A general description for the evaluation of several actinoids data were presented.
\end{abstract}

\begin{keyword}
Actinoid, Nuclear Data, Evaluation, ENDF
\end{keyword}

\begin{pacs}
24.10.-i, 25.85.Ec, 29.87.+g
\end{pacs}


\begin{multicols}{2}

\section{Introduction}

The new generation evaluated nuclear data library, CENDL-3.1, has recently been released
by China Nuclear Data Center\cite{CENDL31}. The evaluation CENDL-3.1 is needed for design
of fission and fusion reactors and for shielding calculations. The actinoid nuclear data
in CENDL-2.1\cite{CENDL21} were re-evaluated taking account of new experimental data, new
standard cross sections from ENDF/B-VI and using a new theoretical model calculation
code. In addition to those nuclides, new 25 nuclides have been evaluated also during the
period between 2000 and 2005. The evaluations
are for all neutron induced reactions with Uranium, Neptunium, Plutonium and Americium in
the mass range A=232-241, 236-239, 236-246 and 240-244, respectively, and cover the
incident neutron energy up to 20 MeV. More detailed actinoid nuclides in CENDL-3.1 are
listed in Table~\ref{tab1}. Comparing with CENDL-2.1, the newly evaluations were ranked
at "New".

In the CENDL-3.1 evaluation, much more efforts were devoted to improve reliability of
nuclide for available new measured data, especially scarce experimental data. Major
aspects of present evaluations are: systematic accumulation, correction and evaluation
of all relevant experimental data; re-normalization of the neutron data to ENDF/B-VI
standard reaction cross sections; assessment of the applicability of several optical
model potentials which obtained before 2000 year for actinoid calculations;
interpretation of the experimental results in terms of nuclear theory to allow
interpolation and extrapolation of the data into unmeasured energy regions; and finally,
assembly of the experimental and theoretical results into formal evaluated nuclear data
files that can be processed for use in applied nuclear programs. The report presents a
general description for the evaluation of several actinoids data.

In the following sections, an overview of actinoid nuclear data of CENDL-3.1 in
comparison with measurements and other database is presented.
\begin{center}
\tabcaption{ \label{tab1}  Actinoid nuclides in CENDL-3.1 database.}
\footnotesize
\begin{tabular*}{80mm}{c@{\extracolsep{\fill}}ccc}
\toprule
     & CENDL-2.1 & CENDL-3.1 & New  \\
\hline
U    & 235,238   & 232-241(10) & 232-234,236,237,239-241 \\
Np   & 237       & 236-239(4)  & 236,238,239             \\
Pu   & 239,240   & 236-246(11) & 236-238,241-246         \\
Am   & 241       & 240-244(6)  & 240,242,242m,243,244    \\
\bottomrule
\end{tabular*}
\end{center}

\section{Resonance Parameters}

Thermal fission and capture cross sections for actinoid nuclides were corrected and estimated
averaging measurements with suitable weights according to method, neutron source, sample purity
and so on. The negative and low-lying resonance parameters were modified to reproduce the thermal
cross sections. The results of thermal fission and capture cross sections for some important
actinoid nuclides are shown in Table~\ref{thermal_tab}. The resonance parameters were adopted from
other databases. According to experimental data and benchmark results, modify some resolved
resonance parameters. On the other hand, the resolved resonance parameters (RRP) of $^{236}$Np and
$^{238}$Np, which were taken from G.B.Morogovskij\cite{Morogovskij02} evaluation results, and the
unresolved resonance parameters (URP) were obtained for $^{238}$Np according to the resolved
resonance parameters. The unresolved resonance parameters were determined so as to smoothly
connect with the evaluated cross sections.
\begin{center}
\tabcaption{ \label{thermal_tab}  Fission and capture thermal cross sections(b).}
\footnotesize
\begin{tabular*}{80mm}{c@{\extracolsep{\fill}}cc}
\toprule
 Nuclide & Fission & Capture  \\
\hline
$^{233}$U    & 7.67654E+01   & 7.52080E+01 \\
$^{234}$U    & 2.98537E-01   & 9.97503E+01 \\
$^{237}$Np   & 2.02034E-02   & 1.60369E+02 \\
$^{238}$Pu   & 1.70116E+01   & 5.61082E+02 \\
$^{240}$Pu   & 6.40123E-02   & 2.87387E+02 \\
$^{241}$Pu   & 1.01199E+03   & 3.61525E+02 \\
$^{242}$Pu   & 1.04233E-03   & 1.91577E+01 \\
$^{241}$Am   & 3.14232E+00   & 6.39448E+02 \\
$^{242}$Am   & 2.09322E+03   & 2.18831E+02 \\
$^{242m}$Am  & 6.39017E+03   & 1.22922E+03 \\
$^{243}$Am   & 6.43805E-02   & 7.67044E+01 \\
\bottomrule
\end{tabular*}
\end{center}

\section{Evaluation Procedure}

There is a significant amount of measurements with different method, facility and neutron source
available for neutron reactions on several of the actinoids in present study. We obtained almost
experimental data from the EXFOR/CINDA database at the Nuclear Data Section of IAEA, the Data Bank
of the Nuclear Energy Agency in Paris, INIS database and relevant periodical literatures. Much of
those experimental data more or less exit discrepancy each other for the same neutron reaction.
For cross sections evaluation, mainly effort was concentrated on experimental data analysis and
evaluation, which include systematic accumulation, correction and evaluation of all relevant
experimental data, and re-normalization of the neutron data to ENDF/B-VI standard reaction cross
sections etc.

Much of the measurements for fission cross-section experimental data and prompt neutron
multiplicities from fission reaction (nubar) are relevant to "standard" or other accurately
measured reactions. The most of nubar measurements are relative to $^{252}$Cf neutron
multiplicities from spontaneous fission, which is very accurately known as 3.7692 $\pm$ 0.0047
\cite{Nichols07,Nichols08}. In the case of fission cross sections, the experimental data are
frequently relative to the fission cross section of $^{235,238}$U or $^{239}$Pu. In present
evaluations, all the measurements of fission cross sections ratios which were evaluated and
corrected before convert them to absolute cross sections using the ENDF/B-VI standard cross
section. And absolute fission cross section experimental data before ENDF/B-VI standard were
normalized to ENDF/B-VI standard. For prompt nubar measurements were corrected to consist with
IAEA recommendation\cite{Nichols07,Nichols08}.

In general, the first step of the evaluation procedure is to accumulate, assess, normalize and
correct the measurements for each isotope. According to fit the measurements for $\sigma_{tot}$,
$\sigma_{non}$ and elastic scattering angular distribution, the neutron optical model parameters
(OMP) are obtained with the APMN code\cite{Shen03}, which is a program for searching automatically
an optimum set of neutron optical model parameters. The theoretical models which are adopted are
mainly coupled-channel optical models (such as ECIS-95\cite{ECIS95} code) and Hauser-Feshbach
statistical plus pre-equilibrium theory (as FUNF\cite{FUNF} code). The sequence usually followed
in the evaluations above resonance region is to optimize agreement of model calculation results to
the evaluated experimental data including cross sections, differential cross sections and double
differential cross section (DDX) by careful model parameter adjustment. After that assembly of the
experimental and theoretical results into formal evaluated nuclear data files as ENDF format.
The final step in the evaluations is to make fine modifications in prompt nubar and other data (generally within
experimental uncertainties) to enhance agreement with simple fast critical benchmark measurements.
\begin{center}
   \includegraphics[scale=0.35]{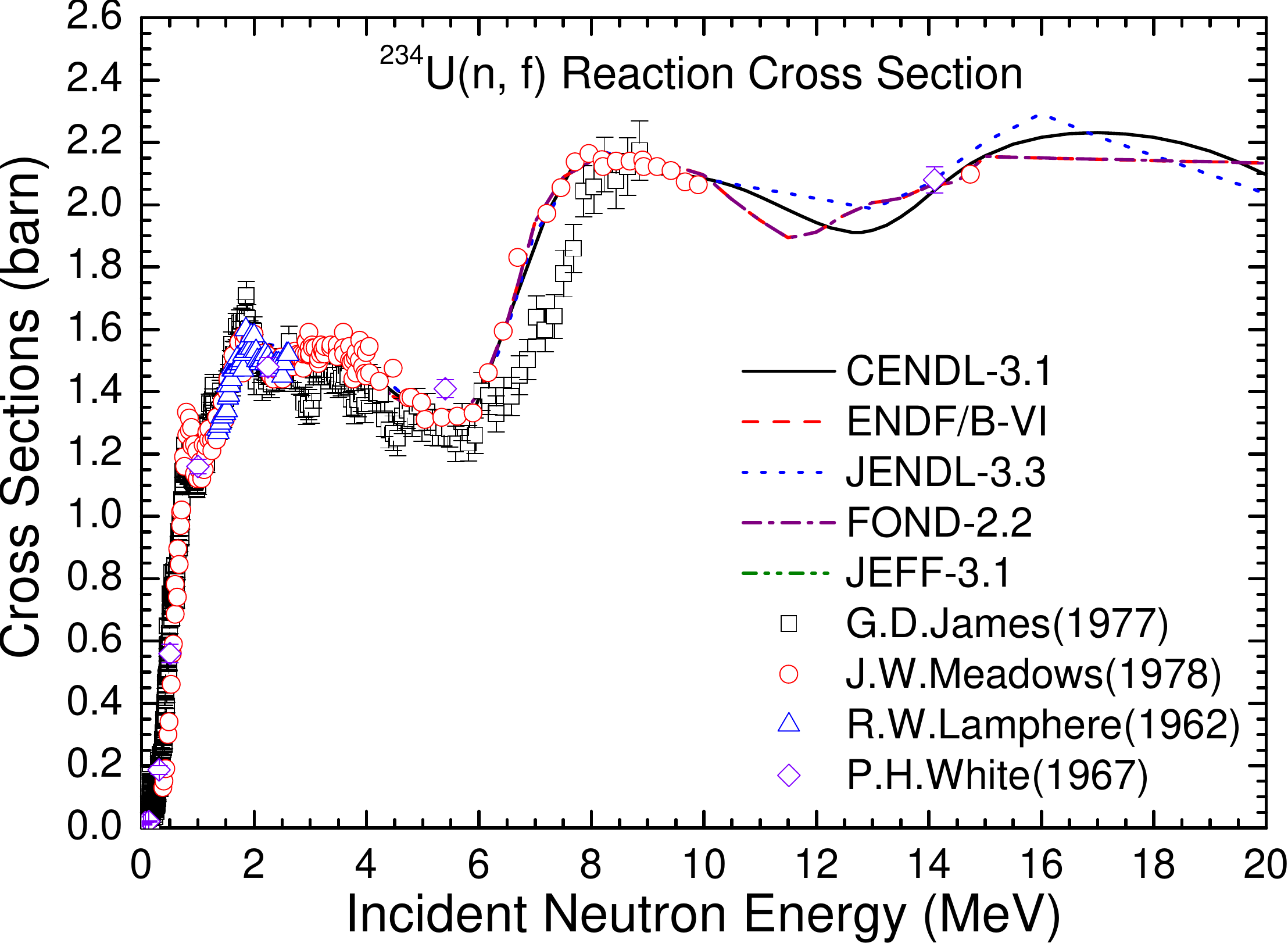}
\figcaption{\label{fig1} Comparison of the evaluated data with the absolute measured data for $^{234}$U(n, f) reaction.}
\end{center}
\begin{center}
   \includegraphics[scale=0.35]{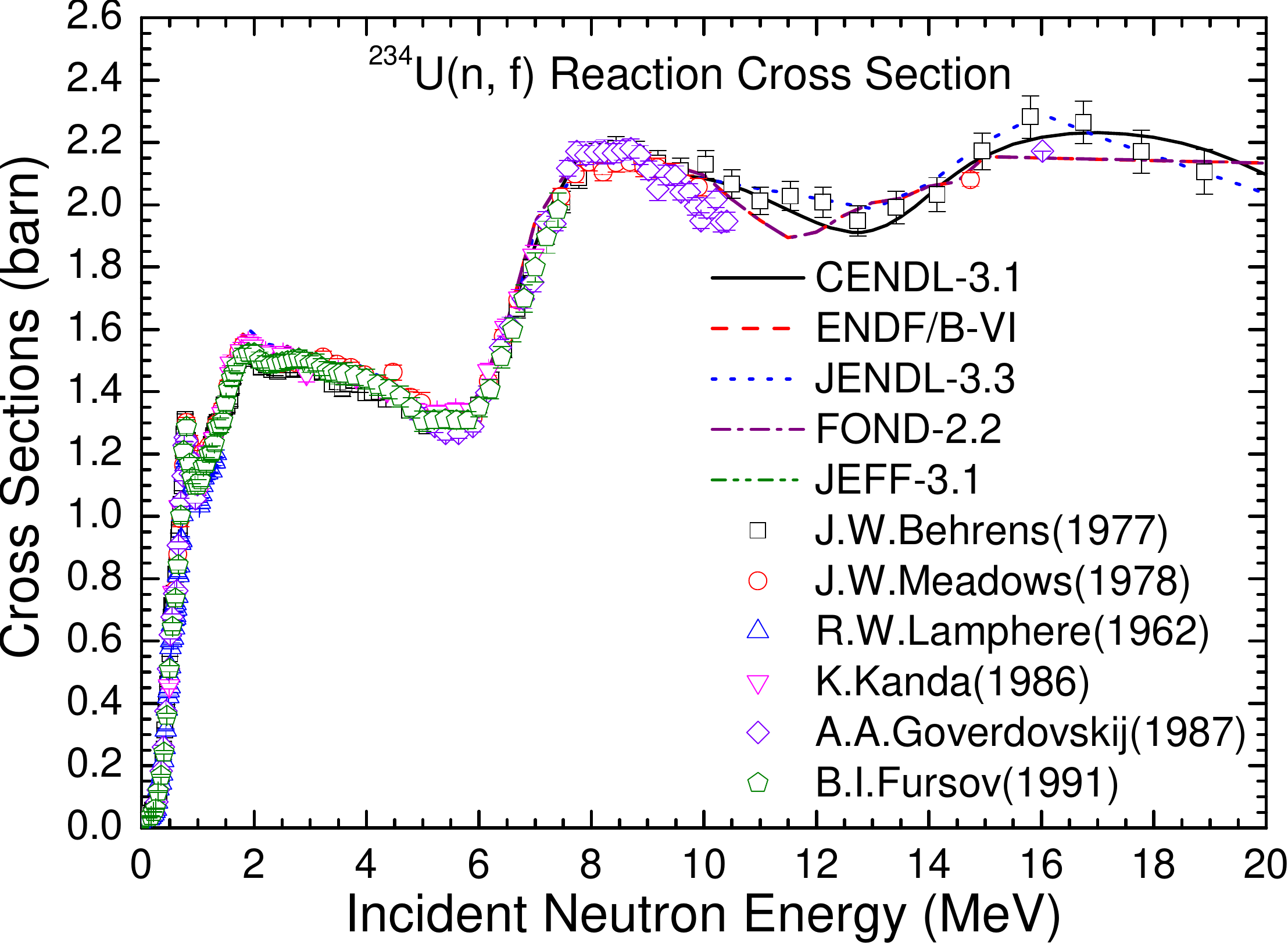}
\figcaption{\label{fig2} Comparison of the evaluated data with the relative measured data for $^{234}$U(n, f) reaction.}
\end{center}

The comparison of the evaluated data with the absolute and the ratio experimental data is shown in
Figs.~\ref{fig1} and ~\ref{fig2}. In Fig.~\ref{fig2}, the ratio measurements were
converted to the absolute one using ENDF/B-VI standard cross sections and compared with the
evaluations. The discrepancy exists in each measurement as shown in Fig.~\ref{fig1}. The ratio
measurement is usually more reliable than the absolute as shown in Fig.~\ref{fig1}. The
original and correction measurements in comparison with each other are shown in
Figs.~\ref{fig3} and \ref{fig4}. There exist discrepancy also for the relative ratio
measurements as shown in Fig.~\ref{fig3}. In consequence, analysis, modification,
normalization and correction is necessary and clarify the discrepancies before using those
experimental data in our evaluation. In Fig.~\ref{fig4} shows the correction relative ratio
measurements in comparison with each other.
%
\begin{center}
   \includegraphics[scale=0.35]{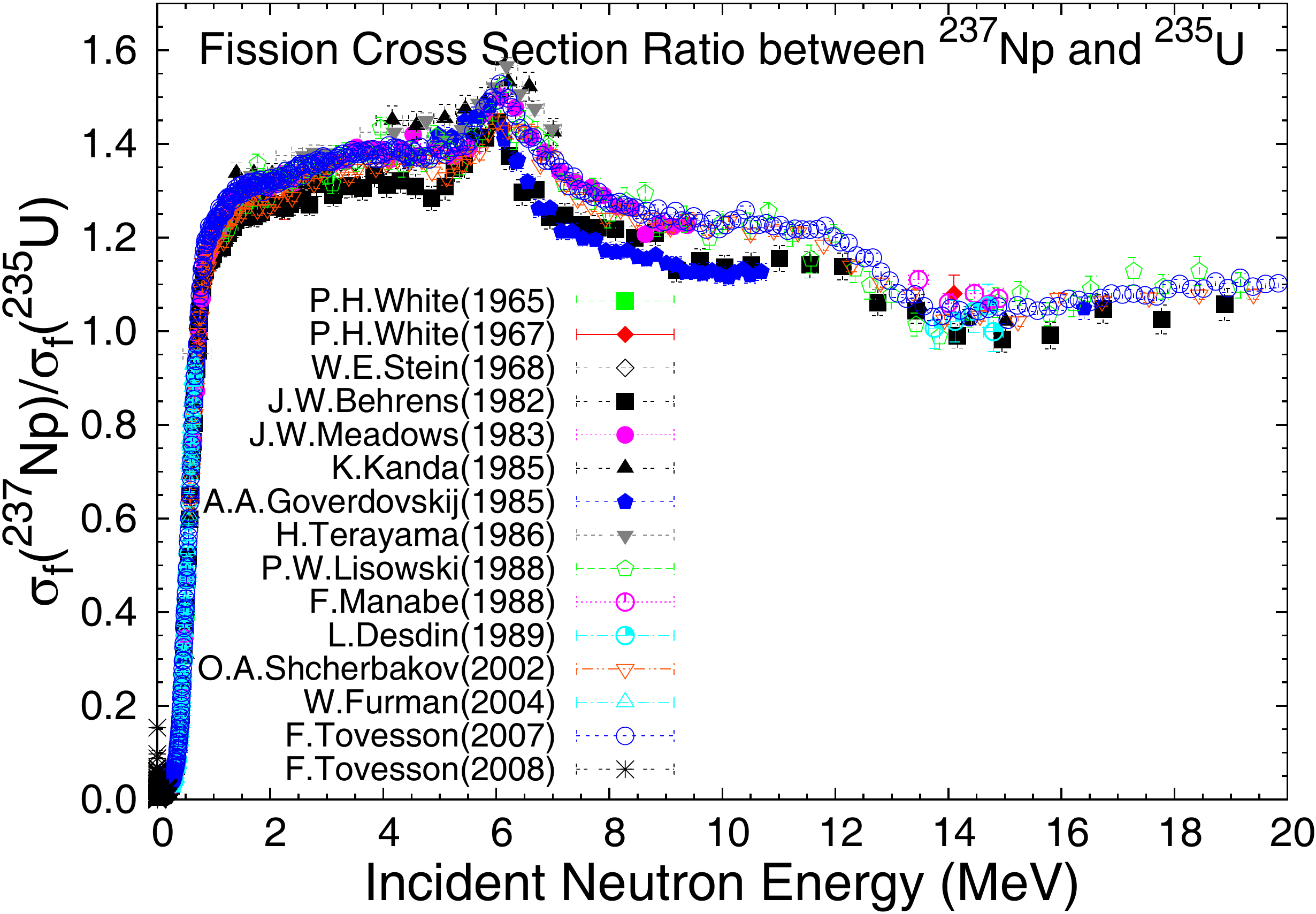}
\figcaption{\label{fig3} Comparison of the original ratio measurements for $^{237}$Np(n, f) reaction.}
\end{center}
\begin{center}
   \includegraphics[scale=0.35]{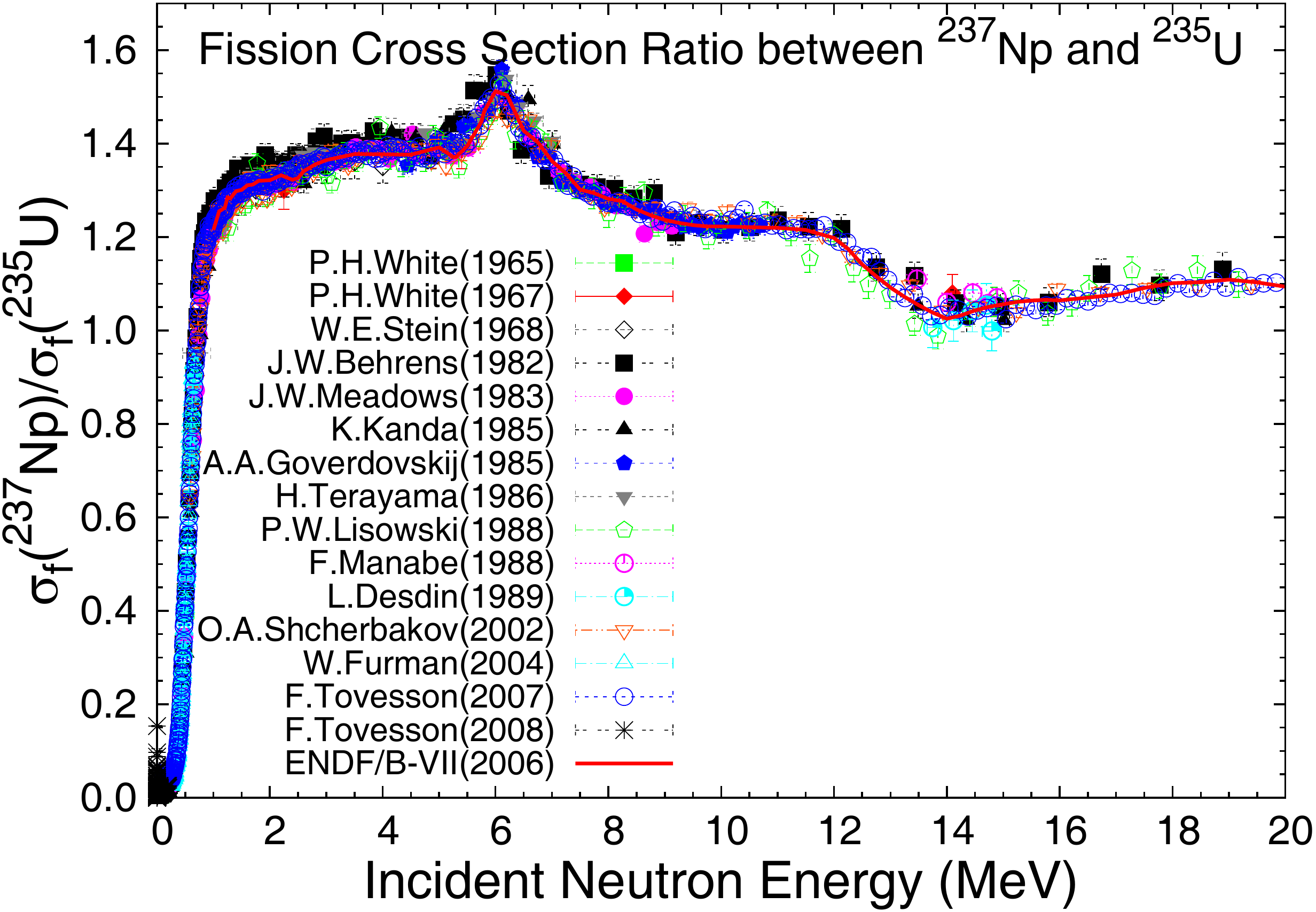}
\figcaption{\label{fig4} Comparison of the corrected ratio measurements for $^{237}$Np(n, f) reaction.}
\end{center}

On the other hand, in order to obtain the reasonable recommended data of actinoid nuclides for
scarce or no experimental data. Through evaluating available experimental data for Uranium,
Neptunium, Plutonium and Americium isotopes, the change tend of some important reaction channel
cross sections such as (n, f), (n, $\gamma$), (n, 2n), (n, 3n), etc. were researched. It was
observed that the reaction channel data depend on the characteristic concerning even-odd (for the
same Z), related fission barrier, level density, pair corrections etc. of actinoid nuclide. A
systematic method to estimate the cross section for (n, f), (n, $\gamma$), (n, 2n), (n, 3n) etc.
reactions were employed for interpolation and$/$or extrapolation suitable scarce data as shown in
Fig.~\ref{fig5} for the comparison of (n, f) reaction cross sections between even mass of Uranium
isotopes. For example, the systematic method was applied to obtain the scarce experimental data
such as $^{246}$Pu(n, f) reaction, which is only exist the experimental data at thermal energy. In
order to recommend the complete neutron data for $^{246}$Pu, the theoretically calculation was
employed based on the change tend to adjust the related fission barriers, level densities and pair
corrections at saddle points for (n, f), (n, n$^\prime$f) and (n, 2nf) phase of the even-even
systematic of Plutonium. The (n, 2n), (n, 3n) etc. reaction cross sections and other qualities
were calculated based on adjusted theoretical parameters in the same way as the even-even
systematic of Plutonium. The systematic calculated fission cross section of $^{246}$Pu is compared
with JENDL-3.3, JEFF-3.1 and ENDF$/$B-VII, that the JEFF-3.1 and ENDF$/$B-VII were taken from
JENDL-3.3 and the plot is shown in Fig. ~\ref{fig6}.
\begin{center}
   \includegraphics[scale=0.35]{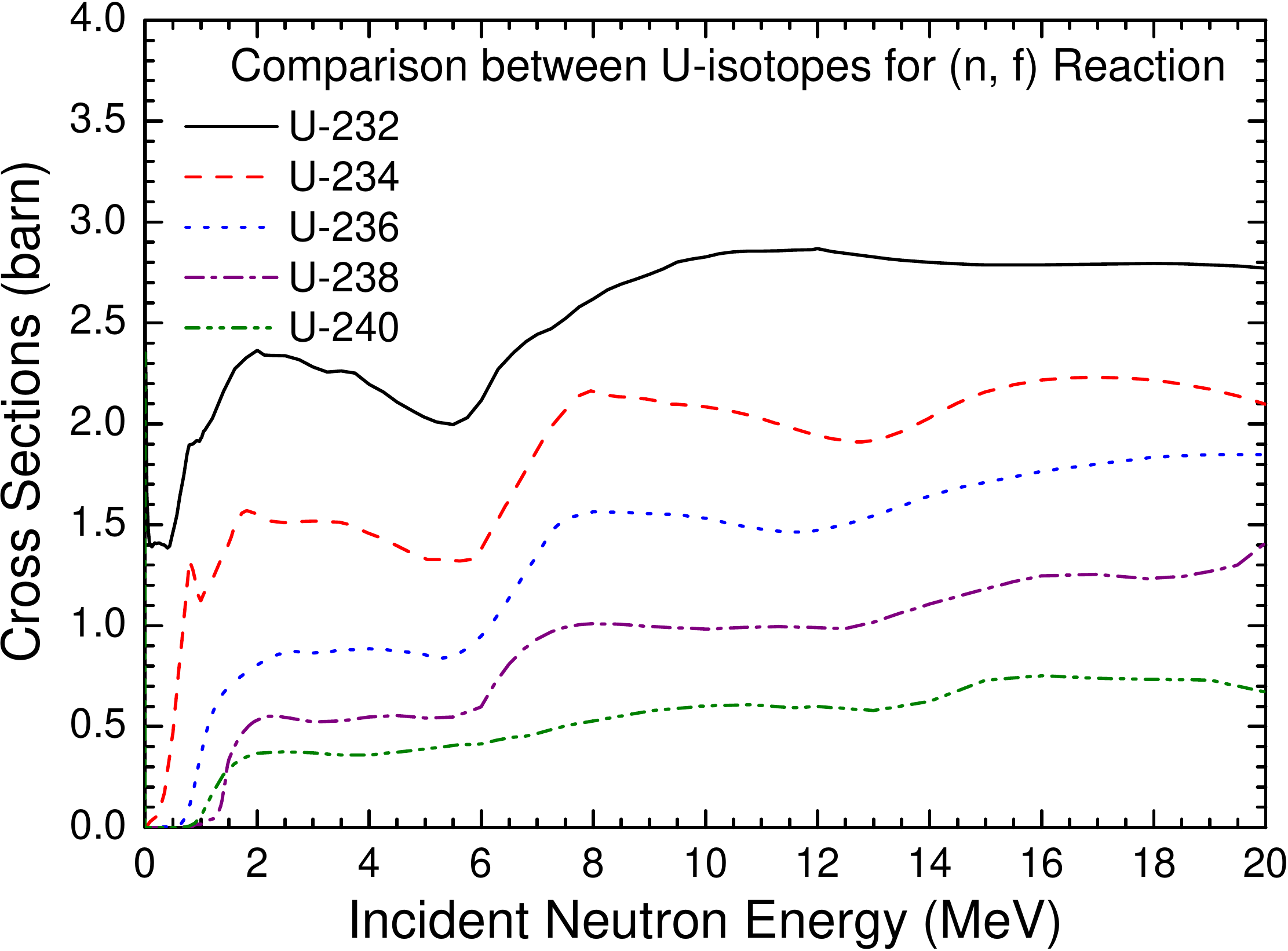}
\figcaption{\label{fig5} Comparison of (n, f) reaction for even mass Uranium isotopes.}
\end{center}
\begin{center}
   \includegraphics[scale=0.35]{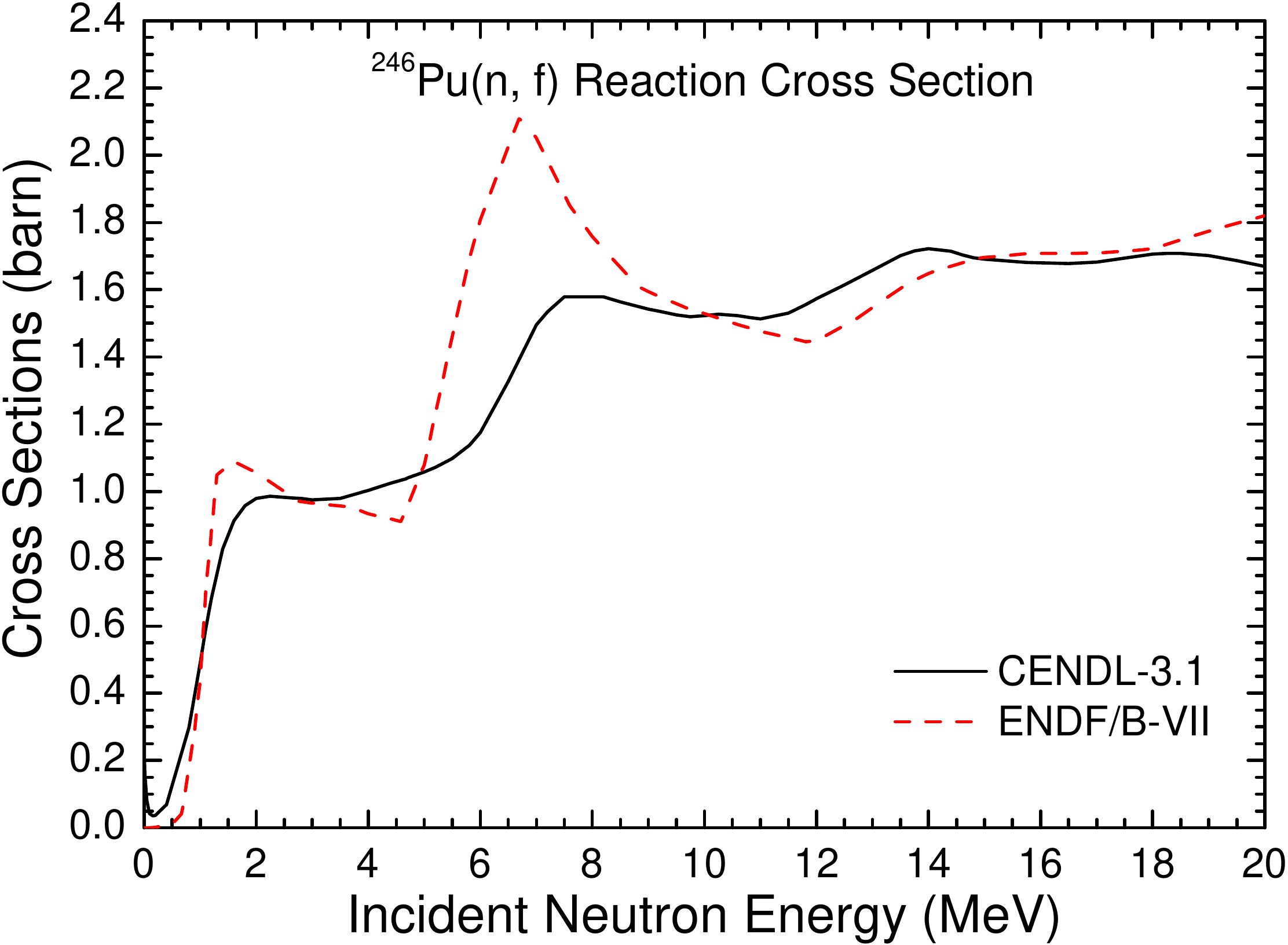}
\figcaption{\label{fig6} Comparison of the evaluated data for $^{246}$Pu(n, f) reaction.}
\end{center}

\section{Conclusion}
Nuclear data of neutron induced reactions were evaluated for 31 actinoid nuclides from U to Am in
the neutron energy range from 10$^{-5}$ eV to 20 MeV during the period between 2000 and 2005. The
evaluated results as a part of CENDL-3.1 were released in December 2009. Actinoid nuclear data in
CENDL-3.1 were widely revised and improved, and 25 new actinoid nuclear data were added in
comparison with CENDL-2.1.


\end{multicols}

\vspace{-1mm}
\centerline{\rule{80mm}{0.1pt}}
\vspace{2mm}

\begin{multicols}{2}

\end{multicols}

\clearpage


\end{document}